\documentclass{ws-procs975x65}

\begin{document}

{\hfill UNF-Astro-3-1-10A}

{\hfill \ } 

\title{GENERALIZED SECOND LAW LIMITS ON THE VARIATION OF FUNDAMENTAL CONSTANTS}

\author{J. H. MACGIBBON}

\address{Department of Physics, University of North Florida,\\
Jacksonville, Florida 32224, United States\\
$^*$E-mail: jmacgibb@unf.edu}

\begin{abstract}
The theoretical maximum time variation in the electronic charge permitted by the Generalized Second Law of Thermodynamics applied to black holes radiating and accreting in the cosmic microwave background matches the measured cosmological variation in the fine structure constant claimed by Webb et al.. Such black holes cannot respond adiabatically to a varying fine structure constant.
\end{abstract}


\bodymatter

\section{Introduction}\label{aba:sec1}
Measurements\cite{W} of absorption in the spectra from distant quasars suggest that the electromagnetic fine-structure constant -- $\alpha = e^2/\hbar c$ where $e$ is the electron charge, $\hbar$ is Planck's constant and $c$ is the speed of light -- may be increasing as the Universe ages, at least in astrophysical environments. These observations motivated Davies {\it et al.} \cite{DDL} to apply the Second Law of Thermodynamics to black holes to derive theoretical limits on $\alpha$ variation. Ref.~\refcite{DDL} and subsequent papers\cite{CV,FT,DK}\,, however, applied the Generalized Second Law\cite{B1} (GSL) incorrectly by investigating the entropy change with respect to $\Delta\alpha$, instead of a time interval $\Delta t >0$. Here we take as our starting point that the net generalized entropy of the black hole system can not decrease {\it over any} $\Delta t >0$, i.e. that $\Delta S_{tot}=\Delta S_{bh} +\Delta S_{env} \ge 0$ where $\Delta S_{bh} $ and $\Delta S_{env} $ are the change in entropy of the black hole and of the ambient radiation and matter, respectively. We ask the question does the GSL applied to a black hole in our present Universe rule out a time variation in $e$ corresponding to the $\alpha$ variation claimed in the quasar measurements, i.e. $de/dt\approx 10^{-23} e$ per second. Below we summarize and update our detailed calculation of Ref.~\refcite{JHM1}. The tightest constraints are derived by considering a charged, non-rotating black hole. In the following we use standard General Relativity and standard QED but extension to theories with additional terms is straightforward. In Section 3 we also discuss adiabaticity.

\section{The GSL applied to Black Holes in the Present Universe}
The entropy\cite{H1} of a black hole is $S_{bh} =kc^{3}A_{bh}/4\hbar G$ where $k$ is the Boltzmann constant, $G$ is the gravitational constant and $A_{bh} = 4\pi G^2\left( M+\sqrt{M^2 - Q^2 /G} \right) ^2 /c^4$ is the area of a non-rotating black hole of mass $M$ and charge $Q$. Thus $dS_{bh}/dt = kc^{3}\left( \partial A_{bh} / \partial t + \partial A_{bh} / \partial e \ de/dt \right)/4\hbar G$ where the first term accounts for black hole accretion and emission and the second term accounts for the cosmological variation in $e$. For quantized charge, $\partial Q / \partial e = Q/e$.

In Case I, $T_{bh}>T_{env}$ where $T_{bh} = 2\hbar G \sqrt{ M^2 - Q^2 /G} / k c A_{bh}$ is the black hole temperature\cite{H1} and $T_{env}$ is the ambient temperature, there is a net radiation loss from the black hole into the environment. If $T_{bh}\lesssim m_e$ (Case IA), the black hole can not discharge via Hawking radiation but looses mass via Hawking radiation at a rate\cite{P1} $dM_H/dt\approx \hbar c^4\beta /G^2 M^2$ where $\beta\approx 3\times 10^{-4}$, so
\begin{equation}
\frac{dS_{bh} }{dt} =\frac{2\pi kG}{\hbar c} \left[1+\sqrt{1-{\frac{Q^{2}}{GM^{2}}}} ^{-1} \right] \left\{\left(M+\sqrt{M^{2} -{\frac{Q^{2}}{G}}} \right)\frac{dM_{H} }{dt} -\frac{Q}{G} \frac{\partial Q}{\partial e} \frac{de}{dt} \right\},
\label{aba:eq1}
\end{equation}
Page has shown\cite{P2} that in this regime $S_{env}$ increases by about 1.62 times the $S_{bh}$ decrease due to Hawking radiation. Thus the GSL would be violated when the first term  in Eq. (\ref{aba:eq1}) is of order the second term, i.e. when $Q$ reaches
\begin{equation} 
Q_{1\approx 2} \approx \left\{\frac{\hbar c^{4} \beta }{GM\left(e^{-1} de/dt\right)} \right\}^{1/2} \left(2-\frac{\hbar c^{4} \beta }{G^{2} M^{3} \left(e^{-1} de/dt\right)} \right)^{1/2} {\rm .} 
\label{aba:eq2}
\end{equation}
Can $Q=Q_{1\approx 2}$ be achieved? If $de/dt\approx 10^{-23}e$ per second, then $Q_{1\approx 2} \lesssim Q_{max} = G^{1/2} M$, the maximal possible black hole charge, for $M\gtrsim 1.8\times 10^{16} {\rm \; g}$. However, if $Q\gtrsim Q_{pp} \approx G^{2} m_{e} ^{2} M^{2} /\hbar ce$, the black hole will quickly discharge\cite{G,ZC,P1} by superradiant \cite{YZ} Schwinger-type\cite{S} $e^{+} e^{-} $ pair-production in the electrostatic field surrounding the hole. Thus superradiant discharge will prevent the black hole charge reaching $Q_{1\approx 2}$ provided that $M\lesssim M_{pp}\sim 5\times 10^{25}\ \rm{g}$ for $de/dt\approx 10^{-23}e$ per second. Remarkably, the mass of a black hole whose temperature is the cosmic microwave background temperature is $M_{cmb}\approx 4.5\times 10^{25}\ \rm{g}$: is this just a coincidence?

If $T_{bh}\gtrsim m_e$ (Case IB), the black hole can discharge via Hawking emission and similarly\cite{JHM1} the entropy increase due to Hawking emission and/or superradiant discharge dominates the entropy decrease due to $de/dt\approx 10^{-23} e$ per second. 

In Case II, $T_{bh}<T_{env}$, there is a net accretion of matter and radiation into the black hole from the environment. A cold black hole in a warm thermal bath will thermally accrete mass at a rate $dM/dt = \hbar c^4 \beta_{env} M^2/G^2 M_{env}^4$ where\cite{P1} $\beta_{env}\sim 10^{-4}$ and $M_{env}$ is the mass of a black hole whose temperature is the ambient temperature. Thus $\Delta S_{tot} \ge 0$ provided $Q < Q'_{1\approx 2}$ where
\begin{equation} 
Q'_{1\approx 2} \approx \left\{\frac{\hbar c^{4} \beta _{env} \left(M/M_{env}\right)^3 }{GM_{env} \left(e^{-1} de/dt\right)} \right\}^{1/2} \left(2-\frac{\hbar c^{4} \beta _{env} }{G^{2} M_{env} M^{2} \left(e^{-1} de/dt\right)} \right)^{1/2} . 
\label{aba:eq3}
\end{equation}
If $de/dt\approx 10^{-23} e$ per second, then $Q'_{1\approx 2}\lesssim Q_{max}$ if $M\lesssim 10^{54}\ \rm{g}$ and $Q'_{1\approx 2}\lesssim Q_{pp}$ if $M\gtrsim 10^{24}\ \rm{g}$. Can a $10^{24}\ \rm{g}\lesssim M\lesssim 10^{54}\ \rm{g}$ black hole achieve $Q\approx Q'_{1\approx 2}$ and so violate the GSL? Two arguments can be used to avoid it. Firstly, a charged black hole will accrete opposite charge fast enough to avoid reaching $Q\approx Q'_{1\approx 2}$. If $\eta_{e^+}$ is the number fraction of positrons in the background, a negatively charged black hole can discharge by positron accretion at a rate $|dQ/dt|_{e^+} \approx 10^{-3} \eta_{e^+} c^3 e T_{env}^3 M^2 / G M_{env}^3$. This will be faster than $(|Q|/e) de/dt$ even when $Q\approx Q'_{1\approx 2}$ provided $\eta_{e^+}\gtrsim 5\times 10^{-17}$ for $de/dt\approx 10^{-23} e$ per second. This condition is met by the positron (and electron) distribution in the present Universe. Secondly\cite{G}\,, a $Q > Q_{acc}\sim 5\times 10^{-22} Q_{max}$ black hole can only gravitationally accrete a like-charged particle if the particle is projected at it with initial velocity and is more likely to lose net charge by accreting a particle of opposite charge; this applies here because $Q'_{1\approx 2} >> Q_{acc}$ for $M\gtrsim 10^{24}\ {\rm g}$.

Combining Cases IA, IB and II, the GSL is not violated by black holes in the present Universe if $e$ is increasing at the rate indicated by the quasar measurements. Because in standard QED $e$ depends on the energy scale of the interaction, the $M_{cmb}$ coincidence we undercovered in Case IA suggests that the measured variation may arise from a coupling between $e$ and the cosmic microwave background.

\section{Do Black Holes Respond Adiabatically to Varying Alpha?}
The entropy of a black hole of general $M$ and $Q$ in the cosmic microwave background cannot be an adiabatic invariant as $\alpha$ increases because the black hole cannot\cite{H3} be in stable equilibrium with its environment and because Hawking emission, thermal accretion and Schwinger-type discharge  -- all of which depend on $\alpha$ -- are fundamentally irreversible processes. Previous papers\cite{DK,FT} investigating black hole adiabaticity with varying $\alpha$ omitted these processes, derived their main results by taking derivatives with respect to $\alpha$ instead of time, and were based on speculative extension models (dilation black holes; quantization of $S_{bh}$ (and hence $\alpha$)). If $\alpha$ is varying, there must be a mechanism by which it varies, either in standard QED or in an extension model. If the $\alpha$ variation arises from new physics beyond the Standard Model, the relevant additional terms should be incorporated into the black hole definitions and $S_{env}$ and the GSL used to derive limits on $\alpha$ variation as above\cite{JHM2}. Because our strongest constraint in Section 2 comes from highly-charged $M_{cmb}$ black holes whose size is much greater than atomic lengths, new physics could only modify our calculation if it introduces terms which are significant on large scales.

\end{document}